# The Role of Magnetic Fields on Astrophysical Jets

Elisabete M. de Gouveia Dal Pino

*Instituto de Astronomia, Geofísica e Ciências Atmosféricas, Universidade de São Paulo, Rua do Matão 1226, Cidade Universitária, CEP 05508-900, São Paulo, SP, Brazil*
*e-mail: dalpino@astro.iag.usp.br*

**Abstract.** Highly collimated supersonic jets and less collimated outflows are observed to emerge from a wide variety of astrophysical objects. They are seen in young stellar objects (YSOs), proto-planetary nebulae, compact objects (such as galactic black holes or microquasars, and X-ray binary stars), and in the nuclei of active galaxies (AGNs). Despite their different physical scales (in size, velocity, and amount of energy transported), they have strong morphological similarities. What is the universal mechanism that can explain their origin? In this lecture, I briefly review the role that magnetic fields seem to play on the formation, structure, and propagation of these jets.
**Keywords**: jets, winds, accretion disks, magnetic fields
**PACS**: 98.58.Fd, 98.62.Nx, 98.62.Mw, 95.30.Qd

## INTRODUCTION

Supersonic jets are ubiquitous phenomena in the universe. Geometrically, they are narrow (small opening angle) conical or cylindrical protrusions that channel mass, momentum, energy and magnetic flux from stellar, galactic and extragalactic objects to the outer medium. They span a large range of luminosities and sizes, from the most powerful examples observed to emerge from the nuclei of active galaxies (or AGNs) to the jets associated to low-mass young stellar objects (YSOs) within our own Galaxy. In the intermediate scales between these two extremes one finds evidences of outflows associated to neutron stars, massive X-ray binary systems (with SS433 being the best example of this class), symbiotic stars, and galactic stellar mass black holes (or microquasars). Also apparently associated to jet phenomena are the gamma ray bursts (GRBs). Discovered nearly three decades ago at cosmological distances, they seem to be the most powerful sources in the Universe (with luminosities $10^{51-53}$ erg/s). There is now some evidence that they are emitted from relativistic jets via Synchrotron emission and, at least those bursts of long duration, are probably associated with supernovae (for a review on GRBs, see Piran 2005, in these Proceedings).

It is worth noting that less collimated supersonic outflows are also observed in the early stages of star formation as bipolar molecular outflows (see, eg., Lery, in these proceedings). They are also often observed to emerge from evolved massive hot stars, like the luminous-blue-variables (LBVs) [29, 60] (see also Owocki 2005, in these proceedings), or from low mass stars in the late stages of stellar evolution, like the proto-planetary nebulae (Garcia-Segura 2005, in these proceedings). A quite similar phenomenon also occurs in a much larger scale in starburst galaxies (i.e., galaxies that suffer episodes of intense star formation), with the formation of gigantic bipolar superwinds that emerge from the galactic disk at high velocities into the intergalactic medium (e.g., Melioli and de Gouveia Dal Pino 2005, in these Proceedings, [41]).

Recent space observations of the Sun have also revealed that even the solar corona is full of jets and flares whose observed spectra and time variability are quite similar to those of cosmic flares and jets (Shibata 2005, these Proceedings).

In this lecture, I will focus mainly on the more collimated type of supersonic outflows, usually called astrophysical jets. As largely stressed in the literature, most of these outflows, despite their different physical scales and power, are morphologically very similar, suggesting a common physical origin (see below). For example, in one extreme, AGN jets have typical sizes $\geq 10^6$ pc, nuclear velocities $\sim c$ (where $c$ is the light speed), and parent sources (which are massive black holes) with masses $10^{6-9} M_\odot$ and luminosities $\sim 10^{43-48}$ erg/s; while in the other extreme, YSO jets have typical sizes $\leq 1$ pc, nuclear velocities $\leq 10^{-3} c$, and emerge from low mass protostars with masses $\sim 1 M_\odot$ and luminosities $(0.1 - 2 \times 10^4) L_\odot$. The jet phenomenon is therefore seen on scales that cover more than seven orders of magnitude. Nonetheless, all the jet classes share common properties. As a rule, they: (i) are highly collimated and in most cases two-sided; (ii) originate in compact objects; (iii) show a chain of more or less regularly spaced emission knots which in some cases move at high speeds away from the central source; (iv) often terminate in emission lobes (with line emission in the case of the YSOs and synchrotron continuum emission in the case of the AGN and microquasar jets), which are believed to be the "working surfaces" where the jets shock against the ambient medium; (v) are associated with magnetic fields whose projected directions are inferred from polarization measurements; and (vi) show evidence of accretion of matter onto the central source via an accretion disk [e.g., 31].

Reviews of the observational and structural properties of the YSO jets can be found in e.g., [52, 53, 49, 23, 17]. Reviews of the properties of the AGN jets and the relativistic galactic jets from X-ray binaries and microquasars can be found in [6, 48, 14, 36, 37, 42, 43, 55, 39, 17]. In this lecture, I will first present a brief review of the currently most accepted mechanism for jet *production* that relies on the presence of large scales magnetic fields and magnetocentrifugal processes (Section 2), then, in Secction 3, I will discuss some recent observational and theoretical results that seem to support this mechanism. In Section 4, I will briefly review recent numerical results of the effects of the magnetic fields on the jet *structure and propagation*, addressing, in particular, YSO jets, and finally I will conclude, in Section 6, with some open questions.

## JET PRODUCTION AND COLLIMATION

Although considerable progress has been made towards understanding the jet structure and propagation, no consensus has been reached concerning the basic mechanism for its origin. However, an universal mechanism of acceleration and collimation that operates in all classes seems to be quite possible.

First, there is observational evidence that almost all systems producing jets contain an accretion disk around the central source. This disk is both a source of energy and provides the required axial symmetry. Second, an examination of all classes of objects which produce jets reveals that the ratio of the observed jet velocity to the escape velocity from the central object, $v_j/v_{escape}$, is of order unity, indicating that jets originate from the center of the accretion disk. In the case of YSOs, there is direct observational evidence (in the form of high-resolution images from the HST) linking jets to the centers of accretion disks (e.g., the jet HH30 [50]). In the case of galactic black hole X-ray transients, the source GRS 1915-105 (a microquasar) provides extra evidence for the connection between the jet and the central part of the disk. Multi-wavelength observations suggest a picture in which the inner disk is episodically accreted while ejecting relativistic plasma which subsequently produces infrared and radio flares by synchrotron emission (see below). Third, these same observations indicate that the jets emerge already collimated very close to the central source.

Therefore, any universal model of jet acceleration and collimation must take these properties into account: (i) the jet has to originate from the disk (or the source) center; (ii) its velocity has to be of the order of the escape (or Keplerian) velocity from the central disk, and (iii) the jet beam must be collimated very close to the central source.

Several mechanisms have been proposed in the literature (see, e.g, [36, 23, 32] for reviews on these mechanisms), but the only one that seems to be able to attend the three conditions above and provide efficient jet acceleration and collimation from the source relies on a rotating accretion disk threaded by a perpendicular large-scale magnetic field [3] (see also the reviews by Shibata 2005, and Cabrit et al. 2005, in these proceedings). The basic idea is that some magnetic flux is in open field lines that form a certain angle with the disk's surface. Ionized material is forced to follow the field lines. Since these lines are anchored into the disk and rotate with it, material is centrifugally accelerated along the field lines like a bead on a wire [59] (see Figure 1). More precisely, when the centrifugal force component along the line exceeds that of gravity, the gas tied to the field line is accelerated outward. This outward magneto-centrifugal acceleration continues up to the Alfvén point [the location where the outflow poloidal speed reaches the Alfvén speed $v_A = B_p/(4\pi\rho)^{1/2}$]. Beyond this point ($\rho v^2 > B_p^2/8\pi$), the inertia of the gas causes it to lag behind the rotation of the field line and the field winds up thus developing a strong toroidal component ($B_\phi$).

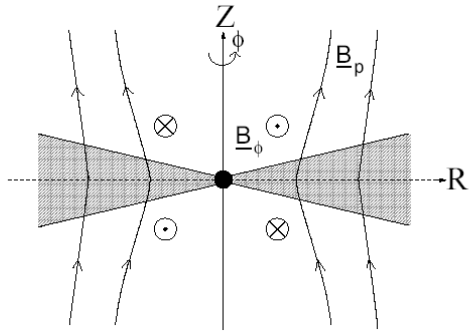

**FIGURE 1.** Schematic view of the magnetocentrifugal disk-driven model. The accretion disk cross-section is shown (shaded) in orbit around the central object. The magnetic field has become advected towards the center along with the accretion flow (see also [1]).

A variant of this magnetocentrifugal *disk-driven* model has been proposed in which the outflow is mainly produced in a tiny region in the interaction zone between the star magnetosphere and the disk system [56, 57]. A schematic diagram of this model is presented in Figure 2. Shu et al. [56] find, for example, that stellar magnetic fields of few kilo-gauss can drive outflows with mass-loss rates of $10^{-6} M_\odot$ yr$^{-1}$ from rapidly accreting YSOs (rotating near breakup) and $10^{-8} M_\odot$ yr$^{-1}$ from slowly accreting (slowly rotating) young stars. This mechanism can accelerate outflows from these systems to the observed velocities of few 100 km yr$^{-1}$ within few stellar radii. In spite of the intrinsic differences that arise from the two classes of magnetocentrifugal models just described, which are in general difficult to distinguish observationally, both classes produce qualitatively similar results (see Cabrit et al. 2005, in these Proceedings, for a detailed discussion on the pros and counters of each class of models within the framework of protostellar jets).

More generally, these mechanisms are believed to be universal in the production of bipolar outflows from a wide range of astrophysical objects. In the case of the AGN jets, there is also evidence for the presence of dusty molecular disks near the central nucleus which gives support to disk-driven or the combined source-disk-driven outflow models.

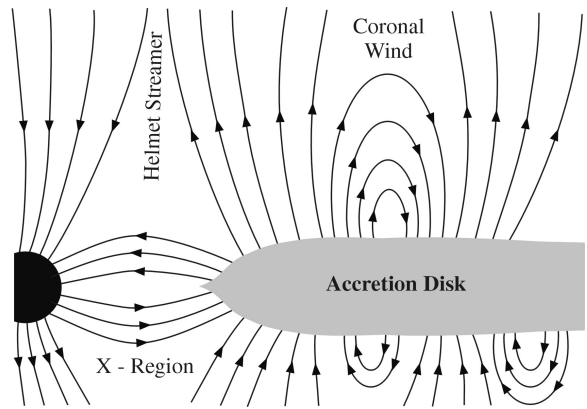

**FIGURE 2.** Schematic representation of the main components of a magnetocentrifugally driven outflow from a *disk-star* system. Gas interior to the X-region ($r < R_X$) diffuses onto field lines that bow inward and is funneled onto the star. Gas at $r > R_X$ diffuses onto field lines that bow outward and launch a magnetocentrifugally driven wind.

Since the outflows are in principle capable of transporting to outside the excess of angular momentum of the accreting matter, as well as, most of the gravitational energy that is liberated, one may attribute the ubiquity of bipolar outflows to the fact that the winds are a *necessary* ingredient in the accretion process in that they carry away the angular momentum that needs to be removed in order to the accretion to proceed.

The collimation process in the magnetocentrifugal models discussed above may occur either by the curvature force ($B_\phi^2/4\pi r$, where *r* is the radial distance from the disk) exerted by the toroidal magnetic field on the outflowing material (collimation by *hoop* stresses), or by the poloidal magnetic flux which increases with the radial distance in the disk (poloidal collimation) if the disk's radius is large relative to that of its central object. The latter process avoids the magnetic kink instability [47] to which hoop stresses collimation is vulnerable.

Numerical MHD models in two and three dimensions ([e.g., 33, 45, 55], see also Shibata 2005, in these proceedings) of the production of magnetized jets are now able to develop jet-like features emerging from the disk central region, but a full self-consistent treatment taking into account all the relevant mechanisms inside the disc, such as the internal and external heating sources, the mass and angular momentum transport, and the physical conditions allowing accretion and ejection is still missing (see below).

# PROBING MAGNETOCENTRIFUGAL MODELS FOR JET PRODUCTION

## With YSO jets

Jets from YSO are probably the best laboratories for cosmic jet investigation, not only because they are, in general, the nearest ones, but also because they are dense enough to produce emission lines from shocks that in turn provide direct information on their physical conditions and kinematics [52,53]. The magnetocentrifugal models described above for jet launching can thus be constrained by the existence of observational data specially from YSO jets.

Within the magnetocentrifugal scenario, if the jet is launched from a rotating accretion disc along with the magnetic field lines that are anchored into the disk-star system, then it should be also rotating, at least near the base. Motivated by this possibility, Bacciotti et al. [2] and Coffey et al. [12] have recently performed high-resolution HST observations of the base of few YSO microjets and detected velocity asymmetries across the jet of 10 to 25 km s$^{-1}$ at distances of few 10 AU from the source, which are consistent with a rotation pattern inside the jet beam.

In order to verify whether this possible signature of rotation could be really unique and unmistakable in the jet flow when other important effects, such as precession and pulsation, were also considered, we have carried out fully three-dimensional Smoothed Particle hydro and magnetohydrodynamics numerical simulations of rotating jets, also including the effects of radiative cooling, precession and velocity variability [11]. Using initial conditions and parameters which are particularly suitable for the DG Tau microjet observed by Bacciotti et al. [2], we have obtained radial velocity maps which are in good agreement with the observations thus indicating that the interpretation by those authors that the DG Tau microjet is rotating should be correct. We have also found that a magnetic field of the order of 0.5 mG should be sufficient to collimate the jet against the lateral expansion that is caused by the centrifugal forces [11] (see also Cerqueira and de Gouveia Dal Pino 2005, in these Proceedings).

These numerical simulations have been also able to reproduce the results recently obtained by Pesenti et al. [46] assuming that the DG Tau microjet is launched magnetocentrifugally by a *warm* accretion disk (see also Cabrit et al. 2005, in these

Proceedings). The consistency between both results is an additional support for the magnetocentrifugal mechanism for jet launching.

We notice however, that an alternative interpretation for these velocity asymmetries have been lately proposed by Soker [60], who has suggested that they could result from the interaction of the jet with a twisted-tilted disk. Further calculations and probably also multidimensional numerical simulations are still required in order to examine this hypothesis in more detail.

## With Microquasar Jets

Binary star systems containing stellar-mass black holes emitting X-rays (also denominated BHXRTs) were first detected in the last decade in our galaxy. Some of them produce collimated bipolar radio jets with apparent superluminal velocities. They are, actually, scaled-down versions of the AGN (or quasar) jets, typically extending for ~ 1pc and powered by spinning black holes with masses of up to few tens of that of the Sun. For this reason they have been named microquasars [42,43]. The X-ray source GRS 1915+105, which is at a distance of only 12.5 kpc and probably hosts a 10 solar-mass black hole, offers excellent features for the investigation of black hole accretion and associated jet phenomena. For example, in a compilation of radio and X-ray data of this microquasar, taken during several weeks, Dhawan, Mirabel & Rodriguez [21] have distinguished two main states of the system, a *plateau* and a *flare* state. The plateau state is characterized by a flat radio spectrum coming from a compact jet of a size of few AU. During this phase, the associated (2-12 keV) soft X-ray emission is weak, while the (20-100 keV) hard X-rays are strong. On the other hand, during the flare phase, superluminal blobs are ejected up to thousands of AU and fade after several days. The soft X-rays also flare during this phase and exhibit high variability, while the hard X-rays fade for few days before recovering. It is generally believed that the X-ray emission of galactic black holes and active galactic nuclei arises from the hot gas accreting onto the central source.

During the plateau phase, the observed the nuclear jet and the 30-minute soft X-ray variability of GRS 1915+105 have been explained in terms of periodic evacuation and refilling of the inner disk region on time scales of seconds as a result of thermal viscosity in the accretion disc [5]. In contrast, the large scale superluminal radio flares observed at the 500 AU scales cannot be explained by the same viscous disk instability model, because they eject an order of magnitude more mass than the AU-scale jet and require a much larger evacuation radius well beyond that where the instability is expected to occur.

Recently, different groups have proposed alternative scenarios to explain the variability of the GRS 1915+105 microquasar. Livio, Pringle and King [38] (see also [30]), for example, have suggested that the inner region of the accretion disk would switch between two states. In one of them, the accretion energy would be dissipated locally by thermal viscosity to produce the observed disk luminosity, and in the other the accretion energy would be converted into magnetic energy and emitted in the form of a relativistic jet. They have attributed the transition between the two states to dynamo generation of a global poloidal magnetic field.

Contemporaneously, in the context of the magnetocentrifugal model of Figure 2, we have proposed that the large scale superluminal ejections observed in GRS 1915+105 during the radio flare events could be produced by violent magnetic reconnection episodes in the corona just above the inner edge of the magnetized accretion disk that surrounds the central 10 solar-mass black hole [20] (see the Helmet Streamer zone in Figure 2).

The process occurs when a large-scale magnetic field is established by turbulent dynamo in the inner disk region with a ratio between the gas+radiation and the magnetic pressures $\beta \approx 1$, implying a magnetic field intensity of $\sim 7 \times 10^8$ G. During this process, substantial angular momentum is removed from the disk by the wind generated by the vertical magnetic flux therefore increasing the disk mass accretion to a value near (but below) the critical one ($dM/dt \approx 10^{19}$ g s$^{-1}$).

We find that the magnetic energy released by reconnection in the helmet streamer zone (Figure 2) is

$$W_B \sim 2 \times 10^{39} \text{ erg s}^{-1} \, \beta_1^{0.94} \, \alpha_{0.1}^{-0.33} \, M_{10}^{1.7} \, R_{X,7}^{3.9} \, l_8^{0.69} \, (dM/dt)_{19}^{2.5} \qquad (1)$$

And the reconnection time is very rapid

$$t_{rec} \sim R_X/(v_A \, \xi) \sim 3.3 \times 10^{-4} \text{ s } \, \xi^1 \, R_{X,7} \qquad (2)$$

where $\beta_1$ is the value of $\beta$ in unities of 1, $\alpha_{0.1}$ is the Shakura-Sunyaev [54] viscous coefficient in units of 0.1, $M_{10}$ is the mass of the black hole in units of 10 solar-mass, $R_{X,7}$ is the inner radius of the accretion disk ($R_X$) in unities of $10^7$ cm, $l_8$ is the is the scale height of the helmet streamer in the corona in units of $10^8$ cm, $(dM/dt)_{19}$ is the disc accretion rate in unities of $10^{19}$ g s$^{-1}$, $v_A$ is the Alfvén speed, and $\xi$ is the reconnection efficiency factor.

Part of the energy above will heat the coronal gas ($T_c \leq 5 \times 10^8$ K) that will be able to produce a steep, soft X-ray spectrum with luminosity $L_X \sim 10^{39}$ erg s$^{-1}$, in consistency with the observations. The remaining magnetic energy released will accelerate the particles to relativistic velocities ($v \sim v_A \sim c$) in the reconnection site through a first-order Fermi process. We find that this produces a steep power-law electron distribution $N(E) \propto E^{-5/2}$ and a corresponding synchrotron radio power-law spectrum with spectral index ($S_\nu \propto \nu^{-0.75}$) which is compatible with the one observed during the flares [20].

## EFFECTS OF MAGNETIC FIELDS ON THE JET STRUCTUTRE AND PROPAGATION

As we have seen above, an important issue in the investigation of the astrophysical jets, is the requirement of the presence of magnetic fields to explain their origin and collimation through magneto-centrifugal forces associated either with the accretion disk that surrounds the central source, or with the disk-source boundary, but what about the effects of the magnetic fields on the structure and propagation of these jets through the environment?

Its importance in the case of the jets produced by active galactic nuclei has long been recognized, actually for more than three decades (see e.g. [6, 39] for reviews). In the case of the YSO jets, more recent polarization measurements [50] have evidenced magnetic field strengths $B\sim 1$ G in the outflow of T Tau S at a distance of few tens of AU from the source, which could imply a $\beta$ plasma ratio, $\beta = p_j/(B^2/8\pi) \sim 10^{-3}$ for a toroidal field configuration (for which the magnetic field decays with the distance $r$ from the jet source as $B \propto r^{-1}$), and $\beta \sim 10^3$, for a longitudinal field (for which the magnetic field decays with the distance from the jet source as $B \propto r^{-2}$), at distances $\sim 0.1$ pc. Since ambipolar diffusion does not seem to be able to significantly dissipate them [25], the figures above suggest that magnetic fields may also play a relevant role on the outer scales of the flow provided that the toroidal component is significant.

In a search for possible signatures of magnetic fields on the large scales of the YSO outflows, several MHD investigations of overdense, radiative cooling jets, have been carried out with the help of multidimensional numerical simulations both in two-dimensions (2-D) [22,25-28,35,44,61-63] and in three-dimensions (3-D) ([7-11], see also [18,19] for reviews).

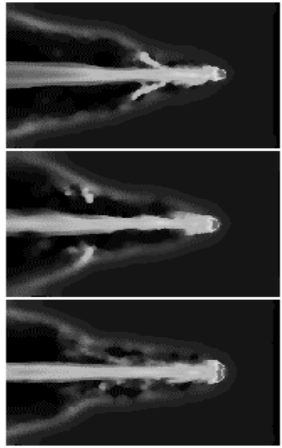

**FIGURE 3.** 3-D SPH simulations of the head of a radiative cooling hydrodynamical jet (top); an MHD jet with initial longitudinal magnetic field configuration (middle); and an MHD jet with initial helical magnetic field configuration (bottom). The initial $\beta$ for the MHD cases is $\sim 1$ and the initial jet velocity is $v_j = 398$ km s$^{-1}$ (see also [7]).

The main results of these studies can be summarized as follows (see also Cerqueira and de Gouveia Dal Pino 2005, in these Proceedings). The effects of magnetic fields are dependent on both the field-geometry and intensity (which, unfortunately, are still poorly determined from observations). The presence of a helical or a toroidal field tends to affect more the characteristics of the fluid, compared to the purely HD calculation, than a longitudinal field (see, e.g., Figure 3 which was extracted from [7]). However, the relative differences that are detected in 2-D simulations involving distinct magnetic field configurations seem to decrease in three-dimensions. In particular, Cerqueira and de Gouveia Dal Pino [9,10] have found that features like the nose cones, i.e., elongated structures that often develop at the jet head in 2-D calculations as a consequence of the pinching of the double shock front by toroidal magnetic fields, are absent in the 3-D models, a result that is consistent with the observations which show no direct evidence for nose cones at the head of protostellar jets. 3-D calculations have revealed that magnetic fields that are initially nearly in equipartition with the gas (i.e., $\beta \sim 1$) tend to affect only the detailed structure behind the shocks at the head and internal knots, mainly for helical and toroidal topologies (Figure 3). In such cases, we have found that the H$\alpha$ emission behind the internal knots can increase by a factor of up to five relative to that of the purely hydrodynamical jet.

This importance of magnetic effects in the bow shocks of protostellar or Herbig-Haro (HH) jets has been recently reinforced by high resolution investigation of individual HH object. For instance, detailed shock diagnosis of the HH 7 object [58] has revealed that the observed H$_2$ emission is due to a C-shock (where heating is provided by ambipolar diffusion) with a magnetic field strength $B \sim 10^{-4}$ G.

As previously stressed [e.g., 18], further 3-D MHD studies are still required as the detailed structure and emission properties of the jets seem to be sensitive to multidimensional effects when magnetic forces are present. Also obvious is the need for further observations and polarization mapping of star formation regions, for a real understanding of their magnetic field structure.

## OPEN QUESTIONS

There is a number of key questions related to the origin and propagation of the jets, and their interaction with the environment. This latter point has actually not been addressed in this lecture, but has been object of ample discussion in the literature (see e.g., [16,17,49], De Colle and Raga 2005, in these Proceedings). Here, I would like just to raise a few questions related to jet production. Among the most relevant, one can mention the problem of the transport of angular momentum in the accretion disc and how this and the accretion affect the disc-jet dynamics. Besides, what are the physical conditions inside the disc that allow for accretion and ejection, and what does make some systems to produce jets and others not? To my knowledge, until the present there is no self-consistent treatment that is able to catch all the aspects of the accretion-ejection mechanism (see the reviews by Bicknell, and King 2005, on accretion discs, in these Proceedings).

There are observational facts (other than those discussed in the previous paragraphs) that hint the possibility that an accretion disk threaded by a magnetic field may be not sufficient, under some circumstances, for the production of powerful jets alone [36, 37]. For example, so far only two of the microquasars (1655-40 and 1915+105) have been observed to produce jets. This suggests that powerful jets may require an additional energy/wind source besides the magnetized accretion disk. In the case of systems with black hole accretors, Livio [36, 37] suggests that the spin of the black hole could provide the extra energy source since rotational energy could be extracted from the spinning black hole [4]. The mechanism itself can be understood if the black hole is compared to a resistor rotating in a magnetic field and generating a potential difference between the hole's pole and equator. Zhang, Cui and Chen [64], in an attempt to determine the spin of the black holes in the microquasars, found that the spin of the sources 1124+68, 2000+25, and LMCX-3 is null, while for 1655-40 and 1915+105, the only microquasars with powerful jets, they found a = 0.93 and 0.998, respectively [where $a=J/(GM^2/c)$ is the dimensionless specific angular momentum of the black hole]. This result is consistent with the suggestion that the spin could be also a necessary ingredient to produce powerful jets in these systems.

Another key question relates to the origin of the magnetic field itself in the jet-accretion disc. Quoting Ferreira [23], "let us face it, we don't know where the magnetic fields come from". They could be either advected from the ambient medium by the infalling material, leading to a flux concentration in the inner disc regions, or could be locally produced by dynamo action in the disc. In the first case, if one neglects dissipation effects, one can estimate the amount advected from the interstellar medium using the law $B \propto n^{1/2}$ [13]. Taking, for a dense ISM cloud, a density $n \approx 1$ cm$^{-3}$ and $B \approx 4 \times 10^{-6}$ G, one obtains magnetic fields of up to $\sim 10^3$ G at a distance of 1 UA [23]. Besides the fact that this is just a rough estimate, as diffusion effects must be important, it is also necessary to compute self-consistently the dynamical effect of this magnetic field along with the energy equation and the ionization state of the matter, a task that may be quite complex (see discussion in [23]). With regard to the dynamo process in accretion discs, numerical simulations are still unable to compute the generation of a large scale magnetic field and most of the models rely on simple physical estimations [e.g., 38]. The treatment of the dynamo is challenging because it depends on the properties of the turbulence generated within the disc. This turbulence, driven by, e.g, magneto-rotational instabilities (see Stone and Gardiner 2005, and Gardiner and Stone 2005, in these proceedings) is generally believed to provide the means not only for an efficient transport, diffusion and heat conduction inside the disc, but also for generation and amplification of small scales magnetic fields. However, a detailed treatment on how these small scale fields turn out into a large scale bipolar field is still missing (see also King 2005, in these proceedings).

## ACKNOWLEDGMENTS

This work has been partially supported by the Brazilian Agencies FAPESP, and CNPq.